\documentclass[a4paper,fleqn,usenatbib]{mnras}

\usepackage{newtxtext,newtxmath}
\usepackage[T1]{fontenc}
\usepackage{ae,aecompl}

\usepackage{graphicx}	% Including figure files
\usepackage{amsmath}	% Advanced maths commands
\usepackage{amssymb}	% Extra maths symbols

\newcommand{\dd}{{\mathrm{d}}}

\title[Solar $p$-mode frequency shifts]{Changes in the sensitivity of solar $p$-mode frequency shifts to activity over three solar cycles}

\author[R. Howe et al.]{
R. Howe,$^{1,2}$\thanks{E-mail: r.howe@bham.ac.uk}
W.J. Chaplin,$^{1,2}$
G.R. Davies,$^{1,2}$
Y. Elsworth,$^{1,2}$
\newauthor
S. Basu,$^3$
A.-M. Broomhall$^4$
\\
$^{1}$School of Physics and Astronomy,
  University of Birmingham, Birmingham, B15 2TT, United
  Kingdom\\ $^{2}$Stellar Astrophysics Centre (SAC), Department of
  Physics and Astronomy, Aarhus University,\\ Ny Munkegade 120, DK-8000
  Aarhus C, Denmark\\ $^3$Department of Astronomy, Yale University, PO
  Box 208101, New Haven, CT 06520-8101, USA,\\
$^4$Institute
  of Advanced Studies, University of Warwick, Coventry CV4 7HS,
  UK\\ $^7$Centre for Fusion, Space, and Astrophysics, Department of
  Physics, University of Warwick, Coventry CV4 7AL, UK}

\date{Accepted XXX. Received YYY; in original form ZZZ}

\pubyear{2018}

\begin{document}
\label{firstpage}
\pagerange{\pageref{firstpage}--\pageref{lastpage}}
\maketitle

% Abstract of the paper
\begin{abstract}
Low-degree solar $p$-mode observations from the long-lived Birmingham Solar Oscillations Network (BiSON) stretch back further than any other single helioseismic data set. Results from BiSON have suggested that the response of the mode frequency  to solar activity levels may be different in different cycles. 
In order to check whether such changes can also be seen at higher degrees, we compare the response of medium-degree solar $p$-modes to 
activity levels across three solar cycles using data from Big Bear Solar Observatory (BBSO), Global Oscillation Network Group (GONG), {\em Michelson Doppler Imager} (MDI) and 
{\em Helioseismic and Magnetic Imager} (HMI), by examining the shifts in the mode frequencies and their sensitivity to solar activity levels. We compare these shifts and sensitivities with those from radial modes from BiSON. We find that the medium-degree data show small but significant systematic differences between the cycles, with solar cycle 24 showing a frequency shift about 10 per cent larger than cycle 23 for the same change in activity as determined by the 10.7\,cm radio flux. This may support the idea that there have been changes in the magnetic properties of the shallow subsurface layers of the Sun that have the strongest influence on the frequency shifts.

\end{abstract}

% Select between one and six entries from the list of approved keywords.
% Don't make up new ones.

%%%%%%%%%%%%%%%%%%%%%%%%%%%%%%%%%%%%%%%%%%%%%%%%%%

%%%%%%%%%%%%%%%%% BODY OF PAPER %%%%%%%%%%%%%%%%%%

\section{Introduction}

The activity-related variation of the frequencies of the solar oscillation modes is one of the earliest and most robust findings of helioseismology, first detected by %... ACRIM, Izana, BiSON, Woodard \& Libbrecht]
\citet{1985Natur.318..449W} using solar irradiance data and later confirmed by 
\citet{1989A&A...224..253P} and \cite{1990Natur.345..322E} for ground-based Sun-as-a-star observations and by \citet{1990Natur.345..779L} for resolved-Sun observations from the Big Bear Solar Observatory (BBSO). 

The frequency variations have now been tracked over nearly four solar cycles. Some recent work has focused on subtle long-term changes in the sensitivity of the frequencies to activity. Recently \citet{2017MNRAS.470.1935H} corroborated the finding of \citet{2012ApJ...758...43B} that the lowest frequency low-degree modes from the Birmingham Solar-Oscillations Network (BiSON) show very little correlation with activity over Solar Cycles 23 and 24, although there were distinct solar-cycle variations in Cycle 22.  \citet{2015A&A...578A.137S} also found little correlation for the low-degree modes in Cycles 23 and 24 using data from the Global Oscillations at Low Frequencies (GOLF) instrument. These findings raised the intriguing possibility that there may have been changes in the magnetic properties of the shallow subsurface layers, namely a thinning of the layer in which magnetic features influence the modes. As a consequence of such a thinning, the lower-frequency modes, for which the top of the cavity lies deeper below the surface than for the higher-frequency modes, would be less affected by changes in the magnetic activity level than they were when the layer was thicker. Additionally, the more recent work \citep{2017MNRAS.470.1935H} found a slight increase in the sensitivity of the medium and higher frequencies to changing activity in the Cycle 24 BiSON data; this again might be explained by a compression of the layer in which the changes are taking place as the activity becomes more concentrated in the thinner layer.

These findings raise the question of whether similar changes in the sensitivity of the frequencies to activity can be seen in the medium-degree modes. Unfortunately, while the BiSON observations extend back to the mid 1970s with good coverage from 1992 onwards, continuous medium-degree observations started only around the beginning of Cycle 23. The Global Oscillations Network Group (GONG; \citealt{1996Sci...272.1284H})  began observing in 1995, while the {\em Michelson Doppler Imager} (MDI; \citealt{1995SoPh..162..129S}) onboard the {\em Solar and Heliospheric Observatory} (SOHO) observed from 1996\,--\,2010, with the {\em Helioseismic and Magnetic Imager} (HMI; \citealt{2012SoPh..275..229S}) onboard the {\em Solar Dynamics Observatory} (SDO) continuing the space-based observing programme from 2010 to the present. Between them, these observations cover Cycle 23 (1996\,--\,2008) and most of Cycle 24 (2009 to the present). The work of \citet{1999ApJ...524.1084H} suggests that the frequency shifts from GONG in the rising phase of Cycle 23 were at least broadly on the same scale as earlier sporadic observations summarized by \citet{1990LNP...367..145L}, and in particular the frequencies obtained from the observations at the Big Bear Solar Observatory (BBSO) between 1986 and 1990, in the rising phase and maximum of Cycle 22.

In this Letter we revisit this comparison using data from GONG, MDI and HMI covering Cycles 23 and 24, together with the BBSO data from Cycle 22 and data from BiSON extending back to 1992. We investigate whether there is consistency between the low-degree BiSON observations and the medium-degree observations from 1986 to 2016, and whether the medium-degree data support the idea of \citet{2012ApJ...758...43B} and \citet{2017MNRAS.470.1935H} about the thinning of the magnetic layer.

\begin{figure*}

\includegraphics[width=0.8\linewidth]{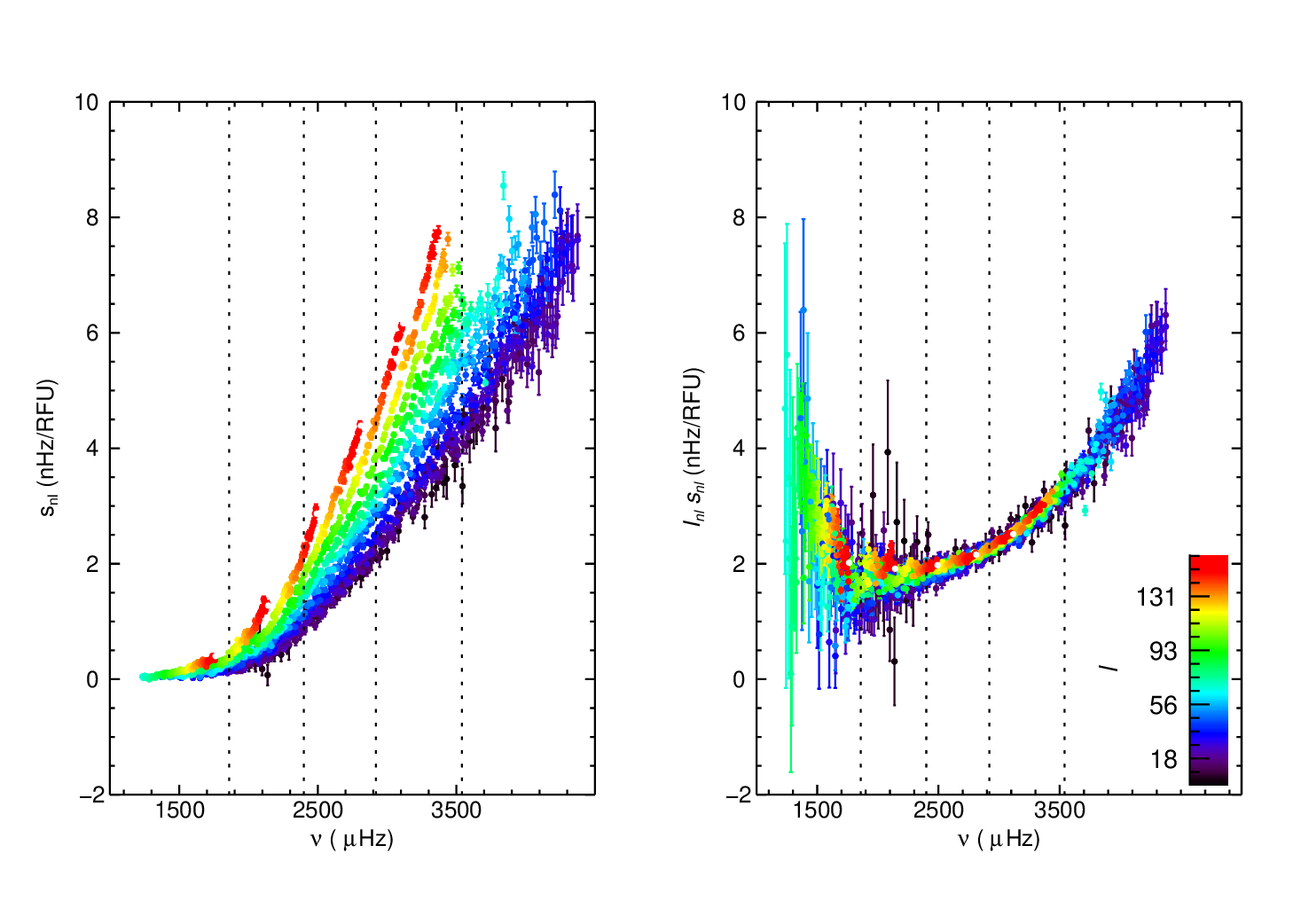}

\caption{Sensitivity $s_{nl}$ (frequency change per unit RF) for individual medium-degree
modes, plotted as a function of 
frequency and colour-coded by the degree $l$, showing the unscaled results (left) and the results scaled by mode inertia (right). The data come from MDI and HMI observations over the period 1996\,--\,2017. The dashed vertical lines indicate the boundaries of the three frequency bands that will be considered in our further analysis.}
\label{fig:fig1}
\end{figure*}

\begin{figure*}

\includegraphics[width=0.8\linewidth]{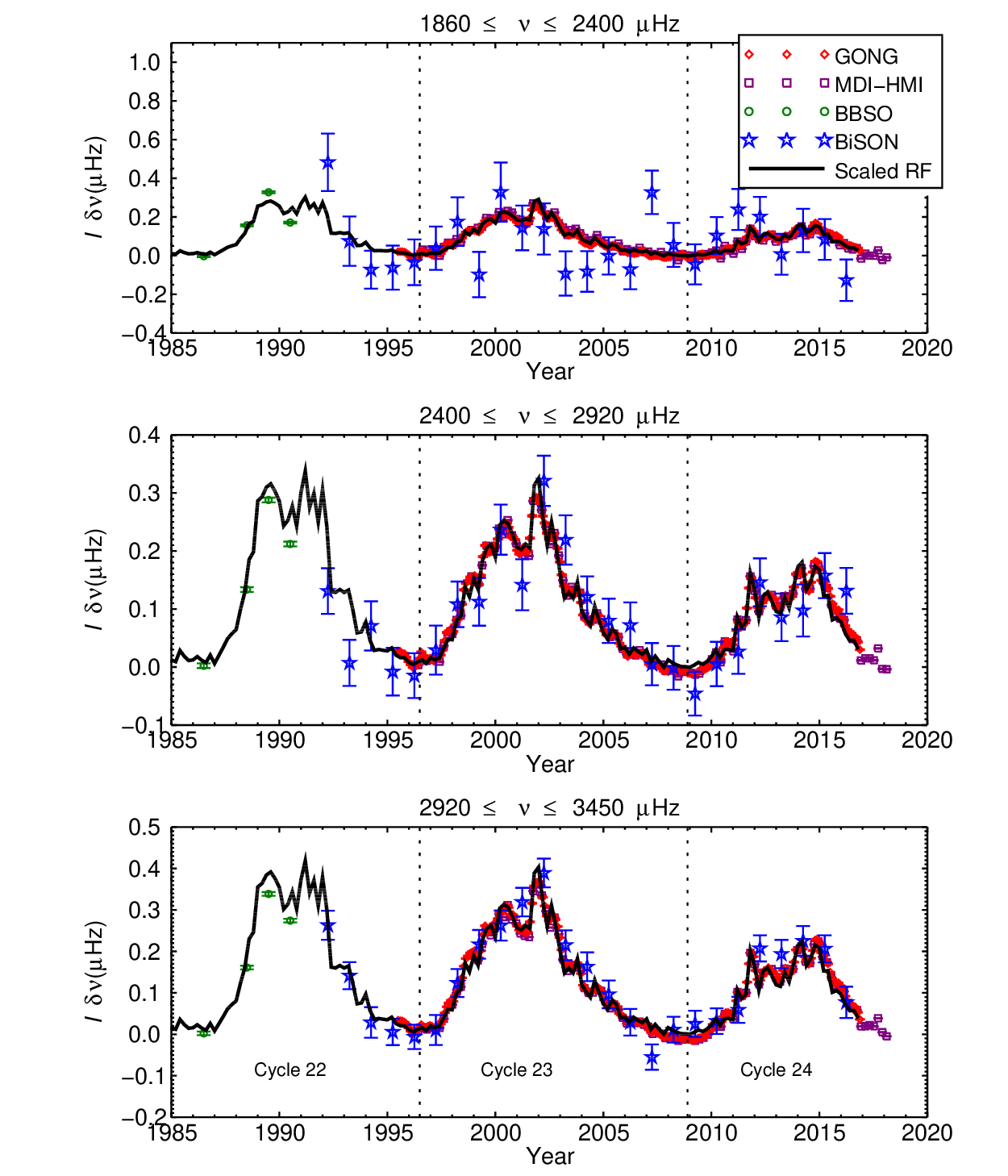}
\caption{Temporal variation of mean inertia-scaled frequency shift for the four data sets in three frequency bands. The dashed line represents the 
10.7\,cm radio flux averaged over 72 days and scaled by a linear 
fit to the MDI--HMI shifts in each frequency band.}

\label{fig:fig0}
\end{figure*}

\begin{figure*}

\includegraphics[width=0.8\linewidth]{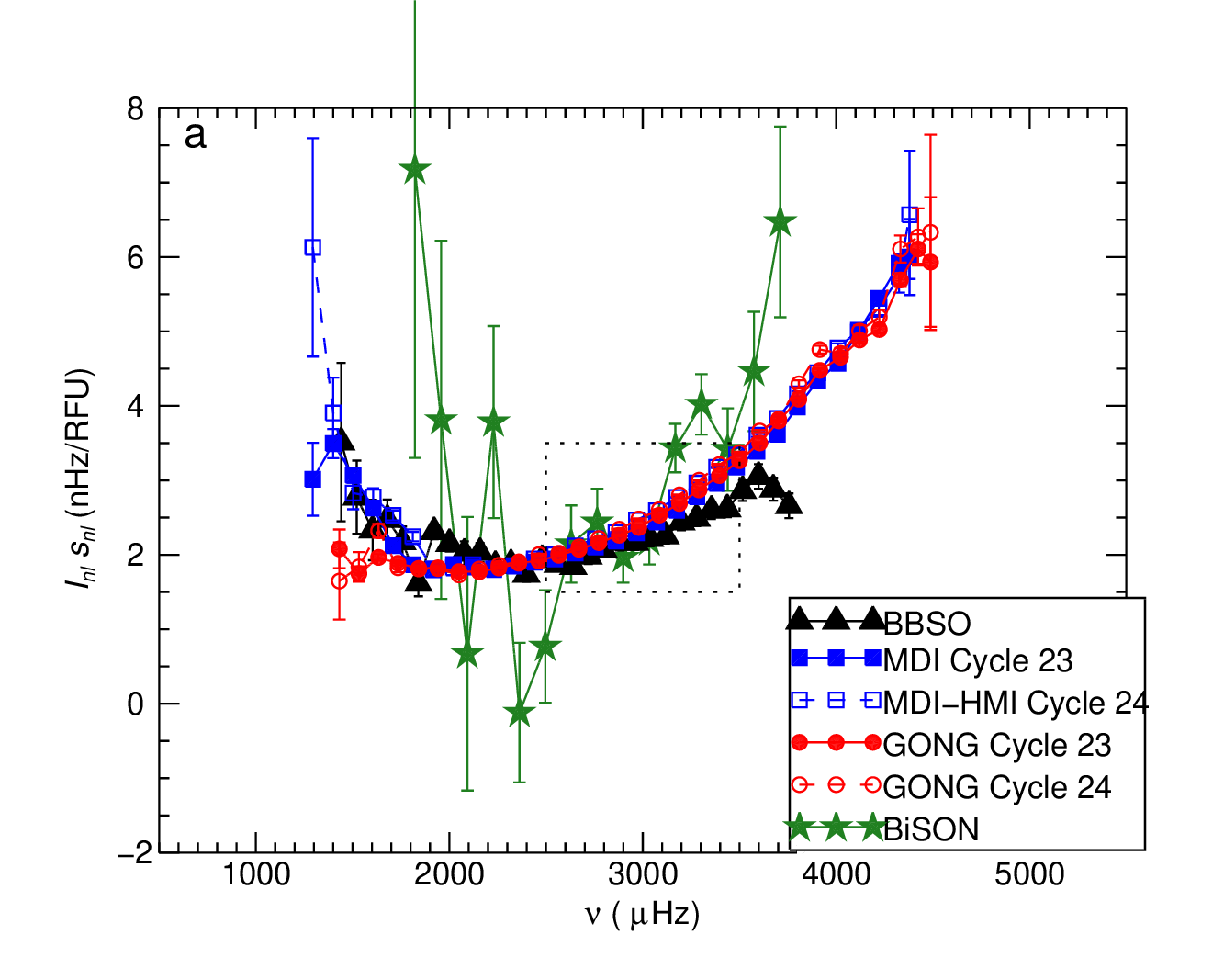}

\includegraphics[width=0.8\linewidth]{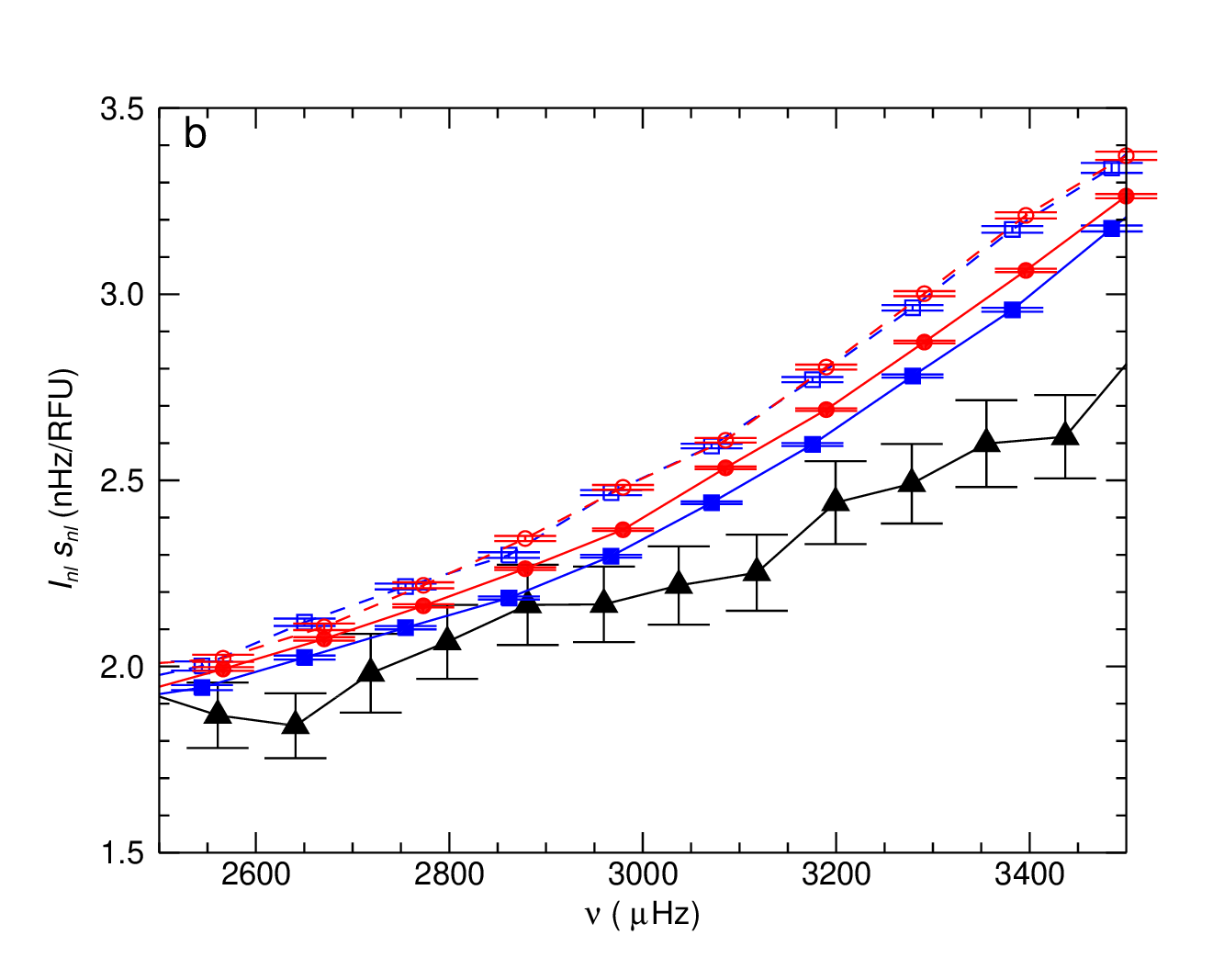}

\caption{Frequency shift per unit RF change scaled by mode inertia to remove the $l$-dependence and binned in frequency, plotted as a function of frequency for BBSO (1986\,--\,1990), GONG (1995\,--\,2017) and MDI--HMI (1996\,--\,2017). The BiSON $l=0$ values are shown for comparison in panel a. The GONG and MDI--HMI data are shown separately for Cycle 23 (1995\,--\,2008) and Cycle 24 (2009\,--\,2017). Panel b shows a detail of the mid-frequency range (indicated by the dotted rectangle in panel a) for the medium-degree data only.}

\label{fig:fig3}
\end{figure*}

\section{Data and analysis}

To investigate the medium-degree mode shifts over the last two solar cycles we used the frequency tables from two projects. The first set of data is from GONG, generated by the GONG analysis pipeline \citep{1996Sci...272.1292H}. The frequencies\footnote{Available from \url{http://gong.nso.edu}} are derived from fitting to 218 overlapping periods of 108 days, with their start times at 36-day intervals. The GONG pipeline attempts to supply a frequency for each 
mode of degree $l$, azimuthal order $m$, and radial order $n$, up to $l=150$. For the purposes of this study we use the central frequency of each $l,n$ multiplet, derived by fitting a polynomial function in $m$ to the individual-$m$ frequencies for each multiplet. The second set of data is based on MDI observations  from 1996\,--\,2011 and HMI observations from 2010 to the present. The pipeline used by the MDI and HMI projects generates multiplet frequencies and polynomial coefficients directly from the fit to the spectrum. We used a total of 104 data sets each derived from 72 days of observations, with the first 71 coming from MDI and the remainder from HMI; we treat the MDI and HMI data together as a continuous set. The tables of frequencies and coefficients that we used came from the most recent improvements to the Stanford pipeline\footnote{Data available from \url{http://jsoc.stanford.edu/lookdata}} at the time of writing; see \citet{2015SoPh..290.3221L} and \citet{Larson2018} for details. 
To compare the two most recent solar cycles with Cycle 22, we also consider the multiplet frequencies from the BBSO observations \citep{1990ApJS...74.1129L,1993ApJ...402L..77W}, which were taken between April and September of the years 1986, 1988, 1989 and 1990. For each medium-degree data set, we used modes with $0 \leq l \leq 150$, and $1 \leq n \leq 30$.

Finally, in order to compare the medium-degree data with the Sun-as-a-star observations and check whether the results are consistent with those of \citet{2017MNRAS.470.1935H}, we use frequencies from the Birmingham Solar Oscillations Network (BiSON; \citealt{1996SoPh..168....1C,2016SoPh..291....1H}) derived from 1-year time periods starting in 1992. These frequencies are from the same analysis used by \citet{2017MNRAS.470.1935H}.
To avoid problems with the latitudinal weighting of Sun-as-a-star observations, which causes the frequency measurements to be most sensitive to the sectoral ($m=l$) modes \citep{2004MNRAS.352.1102C,2017MNRAS.464.4777H}, we used only $l=0$ modes for this analysis. This does result in larger uncertainties compared with those in \citet{2017MNRAS.470.1935H} where we used $0\leq l \leq 3$.

As a proxy of global magnetic activity we used the 10.7\,cm solar radio flux \citep{2013SpWea..11..394T}
\footnote{available from the National Research Council of Canada at \url{ftp://ftp.geolab.nrcan.gc.ca/data/solar_flux/daily_flux_values/}}, 
averaged over the period corresponding to each frequency set. This proxy has been found \citep[see, e.g.][]{2007ApJ...659.1749C} to be one of the proxies that is best correlated with global frequency variations, and it also has the advantage of being available for the whole of the period covered by this analysis.

For the data set from each of the projects (BiSON, GONG, MDI--HMI, BBSO), we compared only those modes for which a frequency was successfully determined in at least 70 per cent of the periods analysed; this choice is made to avoid unduly restricting the common mode set while reducing any bias due to missing modes.
In order to determine a frequency shift, we need a reference frequency for each mode. Some studies use a temporal average, which has the disadvantage that it changes as more data are added or if the data cover different epochs: another possibility is to use the frequencies from a single epoch, but this results in larger uncertainties in the shifts because the uncertainty in the reference frequency is of a similar size to that in the frequencies with which it is being compared.
To avoid these issues, we constructed a reference set of frequencies for each data source by performing a linear least-squares fit of the frequencies of each mode to the corresponding values of the difference of the 10.7\,cm radio flux
 from its lowest value in the data series and taking the intercept of this fit as the `minimum-activity' frequency relative to which all of the shifts were considered. The reference frequency for the same mode may be slightly different for different projects due to the different algorithms used to derive the mode parameters. The slope of such a fit, which we call the sensitivity, $s_{nl}$, 
gives a measure of the response of the mode frequency to the activity index.  
In previous work we have also referred to this quantity as $d\nu/dRF$.

\section{Results}

In Figure~\ref{fig:fig1} we show the $s_{nl}$ values for each mode over the whole MDI--HMI data set, together with the same values scaled by the  mode inertia, $I_{nl}$, from Model S of \citet{1996Sci...272.1286C}. This scaling was originally used for frequency shifts by  \citet{1990Natur.345..779L}. The mode inertia (also known as $E_{nl}$) was defined by \citet{1986ASIC..169...23C} as equal to 
\begin{equation}
{E_{nl}}={\frac{\int_0^R{[\xi_r^2(R)+l(l+1)\xi^2_t(R)]\rho r^2 \dd r}}{4\pi M[\xi_r^2(R)+l(l+1)\xi^2_t(R)]}},
\end{equation}
where $\xi_r$ and $\xi_t$ are the radial and tangential components of the displacement, $M$ is the total mass of the Sun, and $R$ is its photospheric radius. The scaling reflects the way that the frequency changes are concentrated in the near-surface layers so that modes with shallower lower turning points (higher degrees $l$) or shallower upper turning points (higher frequencies) are more affected by magnetic activity. The $I_{nl}$ values are normalized to the value at $l=0$ and a frequency of 3\,mHz. As was seen by \citet{2001MNRAS.324..910C} and \citet{2002ApJ...580.1172H}, for example, the scaling removes most of the 
 `ridge' structure  seen in the left-hand panel of Figure~\ref{fig:fig1} in the sensitivity values, whereby modes of each radial order $n$ lie along distinct curves (with the sensitivity increasing with the degree $l$ more steeply at lower orders), collapsing the sensitivity values into one frequency-dependent curve.  The frequency dependence of this `collapsed' sensitivity curve can be parametrized \citep[see][and references therein]{2017MNRAS.464.4777H}, but this is not the focus of our current work.  The downward trend with increasing frequency in the scaled $s_{nl}$ values at frequencies below about 2500\,$\mu$Hz is not well understood, but it is consistent with the results of \citet{2002ApJ...580.1172H}.

Figure~\ref{fig:fig0} shows the temporal variation of the frequency shift for all of the data sets in the same three frequency  bands as those used by \citet{2012ApJ...758...43B} and \citet{2017MNRAS.470.1935H}, to facilitate comparison with that work. The frequency shifts for each mode, relative to the reference frequency obtained as described above, were scaled by the normalized mode inertia and then averaged (with $1/\sigma^2$ weighting) over each frequency band. Also plotted is the RF flux averaged in 72-day intervals, shifted and scaled by the intercept and slope of a linear fit to the MDI--HMI frequency shifts in each band. The agreement between the GONG and MDI--HMI data is so good that the two curves can hardly be distinguished from one another or from the scaled RF, while the BBSO data lie almost on the scaled RF curve -- perhaps slightly below it for the high-frequency band. There is a clear correlation between the medium-degree frequency shifts and the RF index even in the low-frequency band. The BiSON data with their larger uncertainties broadly agree wth the medium-degree values in the two higher-frequency bands, while in the low-frequency band the uncertainties in the BISON $l=0$ data make it harder to see a clear trend, but the general trend and scale of the variation is consistent with the medium-degree results.

Finally, in Figure~\ref{fig:fig3} we plot the scaled sensitivity values binned in frequency, dividing the GONG and MDI--HMI data into Cycle 23 (dates before the end of 2008) and Cycle 24 (dates from 2009 onwards). In the zoomed-in view of Figure~\ref{fig:fig3}b we can see a small but significant difference between the scaled sensitivity values for medium-degree modes at a given frequency for the two cycles at medium and higher frequencies, with the Cycle 24 curves being systematically about 10 per cent higher than those for Cycle 23.  This difference is consistent with %what was found by
the result of  \citet{2017MNRAS.470.1935H} for BiSON data, where a slightly higher average sensitivity was seen for the medium- and high-frequency bands in Cycle 24 than in Cycle 23. The sensitivity values for BBSO appear to follow a shallower curve, with a slightly lower sensitivity in the high-frequency band and perhaps a slightly higher value at the low-frequency end, while agreeing well with the GONG and MDI--HMI Cycle 23 values in the middle frequency range. When considering only the $l=0$ modes, the uncertainties on the BiSON data do not permit a useful comparison of the frequency dependence of the sensitivity between the cycles. Note that each BiSON data point represents a single mode, while the medium-degree values combine a few dozen multiplets for each point in this plot; the number of modes averaged to obtain each multiplet frequency also scales with the degree $l$. \citet{2017MNRAS.470.1935H} were able to examine the low-degree frequency variations in more detail by considering the modes with $l\leq 3$.

\section{Discussion and conclusions}

We have compared the response to solar activity levels of medium-degree $p$-mode frequencies from GONG, MDI, and HMI in solar cycles 23 and 24 and from BBSO data in the rising phase of solar cycle 22 with that of the $l=0$ mode from BiSON covering the declining phase of cycle 22 up to the present. The results are broadly consistent.

The medium-degree frequencies show a clear correlation with activity even in the low-frequency band, the scale of which appears to be consistent with the scale of the variation of the BiSON data in cycles 23 and 24, given the uncertainties in the low-degree data. 
The 1992\,--\,1994 BiSON results in the low-frequency band do seem to show a larger scale of frequency variation, suggesting that the frequencies have a higher sensitivity to activity than is seen in the more recent cycles, as reported by \citet{2012ApJ...758...43B}. It is not clear that there is a corresponding higher sensitivity in the low-frequency BBSO data from the rising phase of cycle 22. These results raise the question of whether the low level of correlation with activity for the low-frequency, low-degree  modes, in cycles 23 and 24, seen in this work and also by \citet{2012ApJ...758...43B}, \citet{2015A&A...578A.137S}, and \citet{2017MNRAS.470.1935H} in cycles 23 and 24, may be due to the lower signal-to-noise of these observations and the relatively low activity levels in the more recent cycles.

However, interestingly, the medium-degree data at medium and high frequencies show about a 10 per cent increase in the sensitivity of the frequencies to activity  between cycle 23 and cycle 24.  This change is consistent with the findings of \citet{2017MNRAS.470.1935H} for the low-degree BiSON data and would support the idea that the near-surface magnetic layer that influences the frequency shifts has become more compressed in cycle 24. We look forward to seeing what happens in Cycle 25.

\section*{Acknowledgements}

This work utilizes data obtained by the Global Oscillation Network
Group (GONG) program, managed by the National Solar Observatory, which
is operated by AURA, Inc. under a cooperative agreement with the
National Science Foundation. The data were acquired by instruments
operated by the Big Bear Solar Observatory, High Altitude Observatory,
Learmonth Solar Observatory, Udaipur Solar Observatory, Instituto de
Astrof\'{\i}sica de Canarias, and Cerro Tololo Interamerican
Observatory. SOHO is a project of international cooperation between ESA and NASA. HMI data courtesy of NASA/SDO and the HMI science team. We would like to thank all those who are, or have been, associated
with BiSON, in particular P.~Pall\'e and T.~Roca-Cortes in Tenerife
and E.~Rhodes~Jr. and S.~Pinkerton at Mt.~Wilson.  BiSON is funded by
the Science and Technology Facilities Council (STFC), under grant ST/M00077X/1. 
SJH, GRD, YPE, and RH acknowledge the support of the UK Science and
Technology Facilities Council (STFC). Funding for the Stellar Astrophysics
Centre (SAC) is provided by The Danish National Research Foundation
(Grant DNRF106). RH thanks the National Solar Observatory for computing support. SB acknowledges National Science Foundation (NSF) grant AST-1514676.

%The Acknowledgements section is not numbered. Here you can thank helpful
%colleagues, acknowledge funding agencies, telescopes and facilities used etc.
%Try to keep it short.

%%%%%%%%%%%%%%%%%%%%%%%%%%%%%%%%%%%%%%%%%%%%%%%%%%

%%%%%%%%%%%%%%%%%%%% REFERENCES %%%%%%%%%%%%%%%%%%

% The best way to enter references is to use BibTeX:

\bibliographystyle{mnras}
%\bibliography{medshifts} % if your bibtex file is called example.bib

% Alternatively you could enter them by hand, like this:
% This method is tedious and prone to error if you have lots of references
%\begin{thebibliography}{99}
%\bibitem[\protect\citeauthoryear{Author}{2012}]{Author2012}
%Author A.~N., 2013, Journal of Improbable Astronomy, 1, 1
%\bibitem[\protect\citeauthoryear{Others}{2013}]{Others2013}
%Others S., 2012, Journal of Interesting Stuff, 17, 198
%\end{thebibliography}

%%%%%%%%%%%%%%%%%%%%%%%%%%%%%%%%%%%%%%%%%%%%%%%%%%

%%%%%%%%%%%%%%%%% APPENDICES %%%%%%%%%%%%%%%%%%%%%

%\appendix

%\section{Some extra material}

%If you want to present additional material which would interrupt the flow of the main paper,
%it can be placed in an Appendix which appears after the list of references.

%%%%%%%%%%%%%%%%%%%%%%%%%%%%%%%%%%%%%%%%%%%%%%%%%%

% Don't change these lines
\bsp	% typesetting comment
\label{lastpage}
\end{document}